\documentclass[%
 aip,
 amsmath,amssymb,
reprint, onecolumn 
]{revtex4-1}

\usepackage{graphicx}
\usepackage{dcolumn}
\usepackage{bm}

\usepackage[utf8]{inputenc}
\usepackage[T1]{fontenc}
\usepackage{mathptmx}
\usepackage{etoolbox}

\newcommand{\T}{{\bf T}}
\newcommand{\x}{{\bf x}}
\newcommand{\y}{{\bf y}}
\newcommand{\p}{{\bf p}}
\newcommand{\q}{{\bf q}}
\newcommand{\bv}{{\bf v}}
\newcommand{\lm}{\lambda}
\newcommand{\vre}{\varepsilon}

\makeatletter
\def\@email#1#2{%
 \endgroup
 \patchcmd{\titleblock@produce}
  {\frontmatter@RRAPformat}
  {\frontmatter@RRAPformat{\produce@RRAP{*#1\href{mailto:#2}{#2}}}\frontmatter@RRAPformat}
  {}{}
}%
\makeatother
\begin{document}

\preprint{AIP/123-QED}

\title{Weighted least action principle for Maxwell equations}

\author{Jacob Rubinstein}
\altaffiliation{Author to whom correspondence should be addressed: koby@technion.ac.il}
\affiliation{Department of Mathematics, Technion, Haifa 32000, Israel}
\email{koby@technion.ac.il}

\author{Gershon Wolansky}
\affiliation{Department of Mathematics, Technion, Haifa 32000, Israel}


\begin{abstract}
The Fermat principle of least time determines the path of a single ray of light given its initial and final positions and the medium electromagnetic properties. However, a single ray is not a measurable physical object. We derive here an upgraded variational principle for the geometric optics limit of Maxwell equations in an arbitrary medium, based on measuring the intensity of the wave on two planes. The principle provides the complete Fresnel rays bundle connecting associated points at the first and second planes. One of the applications of the present theory is to use the reciprocity between Fresnel rays and phase normals to determine the phase of an electromagnetic wave from two intensity measurements.

\end{abstract}

\maketitle

\section{Introduction}

The weighted least action principle of geometrical optics for scalar wave equations was introduced in \cite{rw04} as a generalization of the Fermat principle of least time. It was further shown \cite{rw05} that the geometrical optics (semiclassical) limit of many dispersive wave equations is related to the optimal transport paradigm \cite{vil} proposed by Monge in the late 18th century. The general framework of optima (mass) transport problem is as follows: Consider two distributions, or intensities, $I_1$ over a domain $\Omega_1$ on a first plane $z=0$, and $I_2$ over a domain $\Omega_2$ on a second plane $z=h$, such that $\int_{\Omega_1} I_1 = \int_{\Omega_2} I_2$. Let $\T(\x)$ be a mapping from $\Omega_1$ to $\Omega_2$. We say that $\T$ transforms $I_1$ to $I_2$ if  it satisfies
\begin{equation}\label{1} \int_{\Omega_1} I_1(\x)\phi(\T(\x)) = \int_{\Omega_2} I_2(\x) \phi(\x)\end{equation}
 for any function $\phi$ continuous on $\Omega_2$. The class of such mappings $\T$ is denoted ${\cal C}$. Note that  if $\T\in {\cal C}$ is differentiable then (\ref{1}) implies
\begin{equation}
I_1(\x)= I_2(\T(\x)) |J(\T)|, \label{p2}
\end{equation}
where $J$ is the Jacobian of the mapping $\T$. Let $C(\x,\y)$ a real valued function on $\Omega_1\times\Omega_2$ be  a cost function.   The Monge problem is to find a mapping $\bar{\T} \in {\cal C}$ that minimizes
\begin{equation}
\int_{\Omega_1} I_1(\x) C(\x,\T(\x)) \;d\x \label{p4}
\end{equation}
within the class $\T\in{\cal C}$.
In the optics context, the mapping $\bar\T$ is induced by the rays connecting points on $\Omega_1$ to points on $\Omega_2$. In \cite{rw04} it is shown that for a proper selection of the cost function $C$, the minimization problem above provides the full ray mapping between two distributions of light. The specific cost function $C$ is determined by the Hamiltonian structure of the wave problem. For example, the optimal mapping $\bar\T$ of the Monge problem for the cost function $C(\x,\T(\x))=n\sqrt{1+|\T-\x|^2}$ provides the ray mapping of the geometrical optics limit of the Helmholtz equation $\Delta u+k^2n^2 u=0$, where $n$ is the refraction index and $k$ is the wave number. Another canonical example is $C(\x,\T(\x))=|\T-\x|^2$ where the optimal $\bar\T$ provides the ray mapping for the paraxial Fresnel equation.

In addition to the theoretical significance of the weighted least action principle, it also serves to solve the phase from intensity problem. The difficult problem of estimating the phase of a wave is solved from two relatively easy intensity measurements, and a single optimization problem \cite{lee06}.

In all the examples listed in \cite{rw05} the rays serve two roles. First, they are directly related to the gradient of the phase $s(x)$. Second, the rays are the transport vectors for the propagation of the intensity. For this reason the optimal mapping $\bar{T}$  provides a solution to the phase from intensity problem.
\par
Our goal in this paper  is to extend the theory to more general wave problems where there are {\it two} distinct families of rays. One family consists of the normal to the wavefronts of the wave, and a different family of rays that are tangent to the direction of  the energy propagation. The chief example of such wave problems is Maxwell equations, and we will concentrate on this case. However, we point out that our analysis and conclusions apply to other anisotropic wave problems, such as those arising in elasticity \cite{BA}. The geometrical optics limit of Maxwell equations in an {\em isotropic} medium is the same as the limit of the scalar wave equation, where the two families of rays coincide. Thus, we are interested here in the {\em anisotropic} case.

In the next section we will recall some essential properties of the geometrical optics limit of Maxwell equations. In section 3 we will derive the equivalence of this limit and an associated mass transport problem. This will enable us to solve the phase from intensity problem for Maxwell equations. Examples of media that are common in practice, and where the calculations can be made explicit are presented in section 4.

A comment on our notation: We use $\x.\p,\q,\T$, etc. to denote two dimensional vectors, and $x,p,q$ etc. to denote three dimensional vectors.

\section{The geometrical optics limit of Maxwell equations}

This section follows the exposition of \cite{lune} and \cite{POO}. We consider time harmonic solutions of the Maxwell equations. The geometrical optics limit of these equations includes two families of curves: Fresnel wave normals, denoting here by $p$ and Fresnel rays denoted here $q$. The curves $p$, as their name indicates, are the normals to the wavefronts. Denoting the phase by $s(x)$ we have $p=\nabla s$. The rays $q$ are associated with the Poynting vectors, and are the carriers of energy. Denoting the electromagnetic energy $G$, the transport equation is
\begin{equation}
\nabla \cdot (qG)=0. \label{p6}
\end{equation}
The geometrical optics limit is equipped with a Hamiltonian structure that has an important geometric characterization. Denoting the Hamiltonian $H$ (not to be confused with the magnetic field!), the Hamilton equation is
\begin{equation}
H(\nabla s,x)=H(p,x)=0. \label{p8}
\end{equation}
The surface $H(p,x)=0$  for fixed $x$ is called {\it Fresnel surface of wave normals}. In an isotropic medium the surface is simply the sphere $|p|^2- n^2=0$, where $n$ is the refraction index of the medium. In general anisotropic media the Fresnel surface of normals consists of two nested shells, whose shape is determined by the dielectric $\epsilon$ and induction $\mu$ matrices. In practice most materials are magnetically isotropic, and to simplify the presentation we assume $\mu=1$. Similarly, the $q$ curves are associated with another surface, {\it Fresnel surface of rays} $H^*(q,x)=0$, that is determined by the inverse matrices $\epsilon^{-1},\, \mu^{-1}$. In section 4 below we write explicit expressions for $H$ and $H^*$ in a reference frame where the matrix $\vre$ is diagonal.

Maxwell equations imply an important geometric connection between $p$ and $q$:
\begin{equation}
q=\frac{\nabla_p H}{p \cdot \nabla_p H}. \label{p18}
\end{equation}
This implies $p\cdot q=1$. Except for rare nongeneric instances where $p$ lies at the intersection of the two Fresnel sheets, we can assume that the phase satisfies  one of the two shells. The dual  relation  to (\ref{p18})
\begin{equation}
p=\frac{\nabla_q H^*}{q \cdot \nabla_q H^*}\label{p18b}
\end{equation}
holds as well.

Since the characteristic equations for the differential equation $H(\nabla s,x)=0$ are
\begin{equation}
x_{\tau} = \nabla_p H,\;\; p_{\tau}=-\nabla_x H, \label{p14d}
\end{equation}
we obtain that when $H=H(p)$ then $p(\tau)=p(0)$. Since $p$ is constant along the characteristic, the first equation of (\ref{p14d}) implies that $x(\tau)$ (and the rays $q$ too following (\ref{p18})) are straight lines.

\section{The Weighted Least Action principle}

Consider two planes $(x_1,x_2,x_3=z=0),\; (x_1,x_2,x_3=z=h)$. Since $q$ is the direction of propagation, the impact, or {\it wave action}, of the electromagnetic wave on each plane is given by $a = G q_3$. We thus define the vector
\begin{equation}
\bv=(\frac{q_1}{q_3},\frac{q_2}{q_3}), \label{p19}
\end{equation}
and rewrite the transport equation (\ref{p6}) in the form
\begin{equation}
\frac{\partial a}{\partial z}+ \nabla_{\x} \cdot (\bv a)=0\label{p20}
\end{equation}
subject to \begin{equation}\label{I1I2} a(0,\x)  =I_1(\x) \ \ , \ \ a(h,\x)=I_2(\x) \ . \end{equation}

It is convenient to use (\ref{p8}) to express $p_3$ (=$s_z$) in terms of $\p:=(p_1,p_2)$:
\begin{equation}
p_3=D(\p,\x). \label{p22}
\end{equation}
We can therefore write $\bv$ as
\begin{equation}
\bv=-\nabla_{\p} D(\p,\x). \label{p24}
\end{equation}

Let $\T^*:\Omega_1\rightarrow \Omega_2$ be the ray mapping obtained by integrating the orbit
\begin{equation}
{\bf x}^*_z={\bv}(z,{\bf x}^*),\;\;\;\x^*(0)=\x, \label{veq}
\end{equation}
where $\bv$ is determined by (\ref{p24}), and setting
\begin{equation}
\T^*({\bf x}) = {\bf x}^*(h). \label{p24b}
\end{equation}
Recalling equation (\ref{p20}), we obtain $\T^*\in {\cal C}$.
Equation (\ref{p24}) implies that the Legendre transform of $D(\p,\x)$ is
\begin{equation}
D^*(\bv,\x)=max_\p D(\p,\x)+\p \cdot \bv \label{p26}
\end{equation}
where the equality holds for $\p=\nabla_{\x} s$, i.e.
\begin{equation}
D^*(\bv,\x)=D(\nabla_\x s,\x)+\nabla_\x s\cdot \bv \label{p66}
\end{equation}
Examples of explicit forms of $D^*$ will be given at the next section.

We are now at a position to write down the Weighted Least Action principle for the geometrical optics limit of Maxwell Equations. Consider two points $\x_1 \in \Omega_1,\;\; \x_2 \in \Omega_2$. The action of an orbit $\x(z)$ connecting them is defined as
\begin{equation}
Q(\x_1,\x_2)=\min \int_{0}^{h} D^*(\x_z,\x) dz, \label{p30}
\end{equation}
where the minimization is over all orbits $\x(z)$ connecting the two endpoints. Let the wave action at $\Omega_1$ and $\Omega_2$ be $I_1(\x)$ and $I_2(\x)$ respectively. Let ${\cal C}$ be the family of mapping $\T$ that transform $I_1$ to $I_2$ in the sense of equation (\ref{p2}). Consider now the problem of minimizing the Monge, or weighted action, functional
\begin{equation}
M(T)=\int_{\Omega_1} I_1(\x) Q(\x,\T(\x))d\x, \label{p32}
\end{equation}
over all maps $\T \in {\cal C}$.

\vskip 0.5cm \noindent {\bf Theorem:}
The ray mapping $\T^*$  defined in equation (\ref{p24b}) is a minimizer $\bar\T$ of (\ref{p32}).


\noindent {\bf Proof:} Consider the complete $z$ derivative of the phase function $s$ along {\it any} orbit $\x(z)$:
\begin{equation}
\frac{d s(\x(z),z)}{dz} = p_3  + \nabla_{\bf x} s \cdot \x_z  =D(\nabla_{\bf x}  s,\x)+ \nabla_{\bf x}s \cdot \x_z\leq
 D^*(\x_z,\x). \label{p34}
\end{equation}
where we used $s_z=p_3$ in the first equality, equation (\ref{p22}) in the second equality and equation (\ref{p26}) for the inequality. If we substitute the orbit $\x^*$ defined in equation (\ref{veq}), and recall equation (\ref{p66}), we get an equality in (\ref{p34}):
$$\frac{d s(\x^*(z),z)}{dz} =
D^*(\x^*_z,\x). $$

Integrating this equation from $z=0$ to $z=h$ we obtain:

\begin{equation}
s(\T^*(\x),h)-s(\x,0)= \int_0^h D^*(\x^*_z,\x)dz \geq Q(\x, T^*(\x))\label{p36}
\end{equation}
where last inequality follows from the definition of the action $Q$.
Multiplying both sides by $I_1(\x)$ and integrating we obtain, using the mass transport property (\ref{1}, \ref{p24b}) of $T^*$
\begin{equation}
\int s(\x,h)I_2(\x) - s(\x,0)I_1(\x)=\int \left(\int_0^h D^*(\x^*_z,\x) dz\right)  I_1(\x)d\x \geq \int Q(\x,\T^*(\x))I_1(\x)d\x, \label{p38}
\end{equation}

Now, let $\bar{\T}(\x)$ be the optimal mapping in the theorem, and let $\bar{\x}(z)$ be the orbit connecting $\x$ and $\bar{\T}(x) $ and achieves the optimal $Q(\x,\bar{\T}(\x))$. If we integrate the flow (\ref{p34}) along $\bar\x(z)$ we obtain
\begin{equation}
\int s(\x,h)I_2(\x) - s(\x,0)I_1(\x) \leq
\int \left(\int_0^h D^*(\bar\x_z,\x) dz\right) =
 \int Q(\x,\bar{\T}(\x))I_1(\x)d\x. \label{p40}
\end{equation}
Equations (\ref{p38}) and (\ref{p40}) together imply $\int Q(\x,\bar{\T}(\x))I_1(\x)d\x \geq  \int Q(\x,\T^*(\x))I_1(\x)d\x$, hence $\T^*=\bar \T$ is the minimal map.

\vskip 0.5cm \noindent {\bf Phase from intensity:} We proceed to apply the weighted least action principle to solve the problem of determining the phase of the wave from two measurements of the energy at two planes along a preferred propagation axis. In the scalar wave equation this problem can be solved directly from the weighted least action principle, since the ray directions obtained from the optimal mapping $\bar{\T}$ are directly related to the normals to the wavefronts. In the present anisotropic medium this is no longer the case. We demonstrate the new solution method for the homogenous case where $H=H(p)$, or $D=D(\p)$. As stated in the preceding section, when $D$ depends only on $\p$, the rays are straight lines. Therefore we can write
\begin{equation}
\bv=(\bar{\T}(\x)-\x)/h. \label{p44}
\end{equation}
The optimization problem (\ref{p32}) becomes to find minimum of
\begin{equation}
M(\T)=\int_{\Omega_1} I_1(\x) D^*(\T(\x)-\x)d\x, \label{p46}
\end{equation}
over all maps $\T \in {\cal C}$. Since $D(\p)$ is a concave function, $D^*$ is convex and the minimization problem is well-defined and the minimizer is unique (see, e.g \cite{Stam} sec. 1.3 ). The optimal mapping $\bar{\T}$ can be found by a variety of known numerical schemes (e.g. \cite{Stam}, Ch 6). The optimal mapping $\bar{\T}$ implies, through equation (\ref{p44}), two components for the 3d vector $q$. A third equation is the Fresnel ray surface $H^*(q)=0$. This determines completely the vectors $q(\x)$ in the domain $\Omega_1$. Equation (\ref{p18b}) determines then the vectors $p(\x)$ on $\Omega_1$ and thus the initial phase.

It is important to note that  while the solution of the Monge problem (\ref{p46}) is unique,  the mapping $\T$ realizing the illumination problem may not be so. Indeed,  as  observed in \cite{rw17},  each critical point of the functional $M$ is a legitimate solution.  Moreover,  each sheet of the Fresnel surfaces implies its own solution.

\section{Explicit expression for the action $Q$}

We present a few examples where the action $Q$, and thus the functional $M$, can be computed explicitly. Assume that the dielectric matrix $\vre$ is diagonalized with the third direction $z=x_3$ being the along the normal to the planes $z=0$ and $z=h$.
Using the eigenvectors to determine the reference frame, the matrix $\vre$ is diagonal with elements $(\vre_1,\vre_2,\vre_3)$. Introducing the two auxiliary functions \cite{lune}, \cite{POO}
\begin{eqnarray} \Phi= \frac{1}{2}\left(
(\frac{1}{\vre_2}+\frac{1}{\vre_3})
p_1^2+(\frac{1}{\vre_1}+\frac{1}{\vre_3}) p_2^2+
(\frac{1}{\vre_1}+\frac{1}{\vre_2})
p_3^2 \right), \label{p12} \\
\Psi=(p_1^2+p_2^2+p_3^2)\left(\frac{p_1^2}{\vre_2 \vre_3}+
\frac{p_2^2}{\vre_1 \vre_3}+\frac{p_3^2}{\vre_1 \vre_2}\right),
\label{p14}
\end{eqnarray}
the two sheets of Fresnel surface are given by
\begin{equation}
H=\Phi-1 \pm \sqrt{\Phi^2-\Psi}=0.
\label{p10}
\end{equation}
Observe that the definitions of $\Phi$ nd $\Psi$ imply that both sheets are homogenous of degree two. Therefore  $p \cdot \nabla_p H =2$, and thus the relation (\ref{p18b}) becomes $q=\nabla_p H/2$.

The direct computation of $D^*$ as the Legendre transform of $D(\p,\x)$ is hard. Instead we compute it directly via the Fresnel ray surface equation. For this purpose we notice that the function $D^*$ presented in equation (\ref{p26}) can be written as
\begin{equation}
D^* = p_3+\p \cdot \bv = (p_1,p_2,p_3) \cdot (q_1/q_3, q_2/q_3 , 1) = p \cdot q /q_3. \label{p51}
\end{equation}
Recalling the identity $p \cdot q=1$ the action $Q$ of a ray connecting
$\x_1 \in \Omega_1,\;\; \x_2 \in \Omega_2$ can be written as
\begin{equation}
Q(\x_1,\x_2)=\min \int_{0}^{h} \frac{dz}{q_3(\x_z,\x)}, \label{p53}
\end{equation}

To find $q_3(\bv,\x)$ we write the two sheets of the Fresnel ray surface $H^*(q,x)=0$ in the form
\begin{equation}
q_3^2(\alpha  \pm \sqrt{\alpha^2-\beta})=1, \label{p55}
\end{equation}
where (suppressing the $x$-dependence of the dielectric coefficients)
\begin{eqnarray} \alpha = \frac{1}{2}\left((\vre_2+\vre_3)\bv_1^2+(\vre_1+\vre_3)\bv_2^2+(\vre_1+\vre_2)\right),
 \label{p57} \\
\beta=(\bv_1^2+\bv_2^2+1)\left(\vre_2 \vre_3 \bv_1^2+\vre_1\vre_3 \bv_2^2 +\vre_1\vre_2 \right).
\label{p59}
\end{eqnarray}
For example, in the isotropic case $\vre_i\equiv \vre=n^2$, we ottain
\begin{equation}
Q(\x_1,\x_2)=\min \int_{0}^{h} n \sqrt{1+\x_z^2}dz. \label{p59}
\end{equation}
Another special case is uniaxial material where $\vre_1=\vre_2 \neq \vre_3$. Here one sheet of the Fresnel ray
surface is $q^2=\vre^{-2}$ and $Q$ is as in equation (\ref{p59}). The action for the other sheet is easily found from equation (\ref{p55}): \begin{equation}
Q(\x_1,\x_2)=\min \int_{0}^{h} n \sqrt{1+\lm \x_z^2}dz. \label{p61}
\end{equation}
where $\vre_1=n^2$ and $\lm=\vre_3/\vre_1$.

\section{Conclusions}

We derived a weighted least action principle for the anisotropic Maxwell equations, as well as for other anisotropic dispersive vector-valued wave problems such as the Cauchy equations of linear elasticity. The theory here differs from the earlier derivation of this principle for scalar (or complex) dispersive waves since in the anisotropic case there are two systems of rays. We used the Hamiltonian structure of Maxwell equations to define a Monge functional related to Fresnel rays. Then, the duality between the Fresnel wave normal surface and the Fresnel ray surface was applied, together with the new least action principle, to solve the problem of phase from intensity. We explicitly computed the cost function for the new variational principle for several common situations.

Finally, we comment that Monge formulated the problem for civil engineering applications and in his original paper he used the cost function $C(\x,\T(\x))=|\T-\x|$. It is interesting to notice \cite{ser25} that shortly after Monge published his paper, Dupin proposed to him that the mapping associated with this cost function might be related to optics. However, while Dupin's intuition is remarkable, his suggestion is not valid since as pointed out above, geometrical optics is actually related to different cost functions.


\end{document}